\newcommand{\rem}[1]{}
\documentclass[12pt]{amsart}
\usepackage{amsfonts,amssymb,amsmath,amsthm,mathrsfs}
\usepackage{enumerate}
\usepackage{url}
\usepackage[dvips]{epsfig}
\makeatletter
\@namedef{subjclassname@2010}{%
  \textup{2010} Mathematics Subject Classification}
\makeatother

\newtheorem{thm}{Theorem}[section]
\newtheorem{lem}[thm]{Lemma}
\newtheorem{prop}[thm]{Proposition}

\newtheorem{remark}[thm]{Remark}
\theoremstyle{definition}
\newtheorem{definition}[thm]{Definition}
\numberwithin{equation}{section}

\begin{document}

\baselineskip=17pt
\title[Simple conformally recurrent space-times are pp-waves]{Simple conformally recurrent space-times \\are conformally recurrent pp-waves}

\author[C. A. Mantica]{Carlo Alberto Mantica}
\address{I.I.S. Lagrange, Via L. Modignani 65, 
20161, Milano, Italy}
\email{carloalberto.mantica@libero.it}

\author[L. G. Molinari ]{Luca Guido Molinari}
\address{Physics Department,
Universit\`a degli Studi di Milano \\ and I.N.F.N. sez. Milano,
Via Celoria 16, 20133 Milano, Italy}
\email{luca.molinari@unimi.it}

\date{Nov 2016}

\begin{abstract}
We show that in dimension $n\ge 4$ the class of simple conformally recurrent space-times
coincides with the class of conformally recurrent pp-waves.
\end{abstract}

%\subjclass[2010]{Primary 53B30, 53B50, Secondary 53C80, 83C15}

\maketitle

\section{Introduction}
Riemannian manifolds with recurrent Weyl tensor (conformally recurrent manifolds) were introduced by Adati and Miyazawa 
\cite{Adati} and subsequently extended to pseudo-Riemannian manifolds by Derdzi\'nski \cite{Der80}, 
Roter \cite{Rot82a,Rot82b,Rot82c,Rot87}, Suh and Kwon \cite{SuhKwon}, and to Lorentzian manifolds
(space-times) by  Mc Lenaghan et al. \cite{LenaghLer,McLenTh}, Hall \cite{Hall74,Hall77}, De and Mantica \cite{DeM2014}. In a recent study on conformally recurrent pseudo-Riemannian and Lorentzian manifolds in dimension $n\ge 5$, 
the authors showed that the Ricci tensor has at most two different eigenvalues, with restrictions on the 
metric structure, and gave an explicit representation of the Weyl tensor as a Kulkarni-Nomizu product \cite{ManMol2016,ManS}.

A pseudo-Riemannian manifold of dimension $n\ge 4$ is ``conformally recurrent" 
if the Weyl curvature tensor\footnote{The local components of the
Weyl tensor are \cite{Postn}
\begin{align*}
C_{jkl}{}^m = R_{jkl}{}^m +\tfrac{1}{n-2}(\delta^m_j R_{kl}-\delta_k^m R_{jl} + g_{kl} R_j^m -g_{jl}R_k{}^m) 
-R\,\tfrac{\delta_j^mg_{kl}-\delta_k^m g_{jl}}{(n-1)(n-2)}
\end{align*}
with Ricci tensor $R_{ij}= - R_{kij}{}^k$  and curvature scalar $R=R^k{}_k$.}
is everywhere non-zero and satisfies the condition
\begin{align}
\nabla_i C_{jklm} = \alpha_i C_{jklm}
\end{align}
where $\alpha_i$ is a non-zero 1-form named recurrence vector.\\ 
The natural question arises whether 
the recurrent vector may be locally cancelled by a conformal transformation. This ``simple" case was 
investigated by Roter \cite{Rot82a,Rot87}.
\begin{definition}[Roter]
A conformally recurrent manifold $(\mathscr M,g)$ is ``simple" if the metric is locally conformal to a %non-conformally flat and 
conformally symmetric metric, i.e. at each point $x\in \mathscr M$ there exist a
neighbourhood $U_x$ and a scalar function $\sigma $ on $U_x$ such that $g'_{ij} = g_{ij} e^{2\sigma} $ and 
$\nabla'_i C'_{jklm} = 0$ on $U_x$.
\end{definition}
The main result of this work is the statement that such spaces, with a Lorentzian metric, coincide with 
(conformally recurrent) pp-wave Lorentzian space-times. 

pp-waves are important in general relativity (see for example \cite{LeistNur} and references therein). They 
were studied in $n>4$ by Schimming \cite{Sch74}, and appear in Kaluza-Klein theories and in string theory. 
In the literature (see for example \cite{Ort13}), pp-wave space-times are identified with Brinkmann-waves, 
i.e. space-times equipped with a null 
covariantly constant vector field, $X^iX_i=0$, $\nabla_j X_k=0$ \cite{Bri25}. Here we conform to the more restrictive
definition used in \cite{Sch74,LeistNur,Lei06a,Lei06b}: 
\begin{definition}
A pp-wave is a Brinkmann-wave whose Riemann tensor satisfies the condition
$R_{jk}{}^{pq} R_{pqlm}=0$.
\end{definition}

The layout of the paper is the following. In section 2 we present preliminary material about simple conformally recurrent 
pseudo-Riemannian manifolds and pp-wave space-times. In section 3 we prove that in dimension $n\ge 4$ a conformally 
recurrent pp-wave space-time is a simple conformally recurrent Lorentzian manifold, and establish the converse implication.
We obtain new curvature properties of conformally recurrent pp-waves, such as 
$\nabla^m \nabla_m R_{ij} =0$. 
In Section 4 we show that conformally recurrent pp-wave space-times 
are solutions to the extended theory of gravity with pure radiation source.

Throughout the paper, the manifolds are assumed to be connected, Hausdorff, with  
non-degenerate metric of arbitrary signature, i.e. $n$-dimensional pseudo-Rie\-man\-nian manifolds. Some results are restricted 
to Lorentzian manifolds (space-times) i.e. to metrics of signature $n-2$. %[31].

\section{Preliminary results}
In this section we present preliminary material about conformally recurrent pseudo-Riemannian 
manifolds, simple conformally recurrent manifolds, and pp-wave space-times.\\

A vector field belongs to the Olszak distribution \cite{Olsz93} if
\begin{align} 
X_i C_{jklm}+X_j C_{kilm}+X_k C_{ijlm}=0. \label{Olsz}
\end{align}
The condition was extensively studied in the geometric literature on 
pseudo-Rie\-man\-nian manifolds  (see for example \cite{DefDes,DesGryc,DesHot,DHS}).\\
A contraction of \eqref{Olsz} with $g^{im}$ gives $X^m C_{jklm}=0$. Then, a contraction with $X^i$ gives
$X^iX_i C_{jklm}=0$. If the Weyl tensor is non-zero, the Olszak distribution consists of null
vectors, $X^iX_i=0$.
\begin{lem} \label{Lem3.9}
On a $n$-dimensional Lorentzian manifold with everywhere non-zero Weyl curvature, the Olszak distribution 
is one-dimensional.
\begin{proof}
Suppose that, besides $X_i$, there exists another covector $Y_i$ satisfying \eqref{Olsz}. Then also $X_i+Y_i$ 
belongs to the Olszak distribution, $X^iX_i=0$, $Y^iY_i=0$ and $X_i Y^i=0$. On a Lorentzian manifold, 
two null orthogonal vectors must be collinear, $Y_i=\mu X_i$ \cite{Santos}.
\end{proof}
\end{lem}

The following general identity holds for the Weyl tensor on a pseudo-Riemannian manifold (eq. 3.8 in \cite{Adati}):
\begin{align}
&\nabla_i C_{jkl}{}^m+\nabla_j C_{kil}{}^m +\nabla_k C_{ijl}{}^m =\tfrac{1}{n-3} \nabla_p \big[\delta^m_j  C_{kil}{}^p \label{idweyl}\\
& + \delta^m_k  C_{ijl}{}^p+\delta^m_i  C_{jkl}{}^p+ g_{kl} C_{ji}{}^{mp} +
g_{il} C_{kj}{}^{mp}+g_{jl}  C_{ik}{}^{mp}\big] \nonumber
\end{align}
\begin{prop}[Adati and Myazawa \cite{Adati}] \label{prop_Adati} 
Let ($\mathscr M, g$) be a conformally recurrent pseudo-Riemannian manifold of dimension $n\ge 4$.
Then:\\
1) $\nabla_m C_{jkl}{}^m=0$ if and only if $\alpha_i$ belongs to the Olszak distribution; \\
2) If $\nabla_m C_{jkl}{}^m=0$, then the recurrent covector is null, $\alpha^i \alpha_i = 0$.
\begin{proof}
Obviously $\nabla_m C_{jkl}{}^m=0$ if and only if $\alpha_m C_{jkl}{}^m=0$. The vanishing 
of the right hand side of \eqref{idweyl} and recurrence imply that $\alpha_i$ belongs to the Olszak distribution. 
Conversely, we have $\alpha^i\alpha_i C_{jklm}=0$, i.e. either $\alpha^i\alpha_i=0$ or $C_{jklm}=0$.
\end{proof}
\end{prop}
\begin{thm}[Roter \cite{Rot82a,Rot87}]\label{th2.2}
A pseudo-Riemannian manifold $\mathscr M$ of dimension $n\ge 4$ is simple conformally 
recurrent if and only if:\\
1) $C_{jklm}\neq 0$ everywhere on $\mathscr M$,\\
2) $\nabla_i C_{jklm}=\alpha_i C_{jklm}$, with recurrence 1-form $\alpha_i$ that is locally a gradient,\\  
3) the Ricci tensor is a Codazzi tensor, $\nabla_k R_{jl}= \nabla_j R_{kl}$.
\end{thm}
The contraction of the Codazzi property with the metric tensor gives $\nabla_k R=0$, where $R$ is
the scalar curvature. Moreover, it was shown:
\begin{prop}[Roter \cite{Rot82a}]\label{prop2.3}
On a non-locally symmetric simple conformally recurrent manifold, $R=0$.
\end{prop}
\begin{prop}[Roter \cite{Rot87}, lemma 12]\label{Lem2.6}
If the Ricci tensor of a non-locally symmetric simple conformally recurrent manifold has the form 
$R_{ij}=\pm d_i d_j$, %being $|d|=1$, 
then the following equations hold:
\begin{align}%(2.3) 
d_i C_{jklm}+d_j C_{kilm}+d_k C_{ijlm}=0, \label{(2.3)} \\
d_i R_{jklm}+d_j R_{kilm} + d_k R_{ijlm}=0. \label{(2.4)}
\end{align}
\end{prop}
\begin{prop}[Roter \cite{Rot87}, theorem 1]\label{Prop2.5}
%Every simple conformally recurrent manifold $\mathscr M$ with a metric of index 1 (or $n-1$) is Ricci-recurrent\footnote{the index of a 
%symmetric matrix is the number of negative entries in its diagonal form (see \cite{[Rot87]},page 324)}. \\
Every simple conformally recurrent Lorentzian manifold is Ricci-recurrent, i.e. there exists a non-zero 
covector field $\omega_i$ such that $\nabla_k R_{jl}=\omega_k R_{jl}$.
\end{prop}

%{\quad}\\

Brinkmann and pp-wave space-times arise in the presence of a null covariantly constant vector field. They are
special cases of Lorentzian manifolds with a recurrent null vector field, which have been studied for a long 
time (see for example \cite{Wal49,Bri25,Sch74,Lei06a,Lei06b}). A characterization of the metric through a 
set of canonical coordinates was obtained by Walker \cite{Wal49}. We recall Proposition 1 of \cite{Lei06b}:
\begin{prop}\label{2.7}
Let ($\mathscr M, g$) be a Lorentzian manifold of dimension $n=d+2>2$ with a recurrent null vector field,
$X^k X_k=0$, $\nabla_k X_j = p_kX_j$.\\
1) This is equivalent to the existence of coordinates $(u,x^1,\dots,x^d,v)$, 
in which the metric has the local expression
\begin{align*}  %\label{(2.5)} 
ds^2 = 2dudv+H(u,\vec x,v)du^2+\sum_{\rho =1}^d a_\rho (u,\vec x)du dx^\rho + \sum_{\mu,\nu=1}^d g^*_{\mu\nu} (u,\vec x)dx^\mu dx^\nu,
\end{align*}
where $g^*_{\mu\nu}$ and $a_\rho$ are independent of the coordinate $v$, and $H$ is a smooth function. We refer to 
these coordinates as Walker coordinates.\\
2) $\nabla_k X_j=0$ if and only if $H$ does not depend on $v$. 
We refer to these coordinates as Brinkmann coordinates, and to the manifold as a Brinkmann-wave (space-time) \cite{Bri25}.
\end{prop}
A pp-wave is a Brinkmann-wave with some restrictions. Schimming \cite{Sch74} gave the following coordinate characterization (see also \cite{LeistNur,Lei06a,Lei06b}):
\begin{prop}[Schimming \cite{Sch74}]\label{Lem2.9}
A Lorentzian manifold of dimension $d+2>2$ is a pp-wave if and only if there exist
coordinates $(u,\vec x, v)$ in which the metric has the following local expression:
\begin{align}
ds^2 = 2du\,dv + H(\vec x, u)du^2 + \sum_{\rho =1}^d dx^{\rho 2} \label{2.6metric}
\end{align}
where $H(\vec x,u)$ is a smooth function independent of $v$, usually called the ``potential 
function of the pp-wave".  
\end{prop}
\noindent
The expression of the metric yields the Ricci tensor of a pp-wave:
\begin{align}
R_{kl} = \psi(\vec x,u) X_k X_l, \qquad \psi(\vec x,u) = -\frac{1}{2} \sum_{\rho =1}^d \frac{\partial^2 H}{\partial x^{\rho 2}} 
\label{eq2.10}
\end{align}
where $X_k=\nabla_k u$ is a covariantly constant null vector ($X^kX_k=0$, $\nabla_j X_k=0$)  \cite{Pod09}. It follows
that the scalar curvature is zero, $R=0$.
\begin{remark}
A metric such that $R_{kl}= \psi X_k X_l$ with a null recurrent covector is called a
pure radiation metric with parallel rays, or aligned pure radiation metric \cite{Lei12}.
\end{remark}
\begin{prop}[Schimming \cite{Sch74}, see also \cite{LeistNur,Lei06a,Lei06b}]\label{Lem2.10}
A Lorentzian manifold of dimension $d+2>2$ with covariantly constant null vector field (i.e. a Brinkmann wave) 
is a pp-wave if and only if one of the following conditions is satisfied:
\begin{align}
& X_i R_{jklm} + X_j R_{kilm}+X_k R_{ijlm}=0;  \label{222}\\
& R_{jklm} = X_jX_mD_{kl} - X_jX_lD_{mk}-X_kX_m D_{jl}+X_kX_l D_{jm};\\
& R^p{}_{jk}{}^q  R_{plmq}=\chi X_jX_kX_lX_m;
\end{align}
$D_{ij}$ is a symmetric tensor and $\chi$ is a suitable scalar function.
\end{prop}
A contraction recovers the form of the Ricci tensor of a pp-wave, $R_{kl}=\psi X_k X_l$. 
Leistner and Nurowski showed that in dimension $d+2=4$ 
the conditions are equivalent to $R_{jk}{}^{pq} R_{pqlm}=0$ \cite{LeistNur}.

\begin{remark} \label{rem_2.11}
It is worth noting that for a Ricci tensor of the form \eqref{eq2.10} we have
\begin{align}
X_iC_{jklm}+X_j C_{kilm}+ X_k C_{ijlm} = X_iR_{jklm}+X_j R_{kilm}+ X_k R_{ijlm}.
\end{align}
A contraction with $g^{im}$ and the equality $R_{ij}=\psi X_iX_j$ imply $X^mC_{jklm}=X^m R_{jklm}$.\\
For a pp-wave, by eq.\eqref{222}, we have
$X_iC_{jklm} +X_j C_{kilm}+X_k C_{ijlm}=0$, and so $X^mC_{jklm}=X^mR_{jklm}=0$.
\end{remark}

\section{Simple conformally recurrent Lorentzian manifolds and pp-waves.}
In this section we obtain two important results. First we show, in several steps, that a conformally 
recurrent pp-wave space-time is a simple conformally recurrent space-time. 
Next, we show that a simple conformally recurrent space-time is a pp-wave space-time. 
Finally, some curvature properties of conformally recurrent pp-wave space-times are presented, 
as well as some technical results.\\

\begin{lem} \label{Lem3.1}
On a pp-wave space-time of dimension $n\ge 4$, the Ricci tensor: 
1) is recurrent with a closed recurrence 1-form, 2) it satisfies  $[\nabla_i ,\nabla_j] R_{kl}=0$, 
3) it satisfies 
\begin{align}
\nabla^j\nabla^m C_{jklm} = -\frac{n-3}{n-2} \nabla^2 R_{kl}\label{eq33}
\end{align}
where $\nabla^2= \nabla_k\nabla^k$. In particular, $\nabla^2 R_{kl}=0$ 
if and only if $\nabla^j\nabla^m C_{jklm}=0$.
\begin{proof}
In a coordinate system with \eqref{2.6metric} one has $\nabla_jR_{kl} = (\nabla_j\psi )X_k X_l = (\nabla_j \log |\psi |) R_{kl}$. 
From this we infer $[\nabla_i,\nabla_j]R_{kl}=0$, i.e. a pp-wave space-time is Ricci semi-sym\-me\-tric \cite{Rotersemi}.\\
The divergence of the Weyl tensor has the general expression
\begin{align}  %\label{(3.1)}
\nabla_m C_{jkl}{}^m = \tfrac{n-3}{n-2} (\nabla_k R_{jl}-\nabla_j R_{kl}) +\tfrac{n-3}{2(n-1)(n-2)}
(g_{kl}\nabla_j R  - g_{jl}\nabla_k R) \label{divC}
\end{align}
A further covariant divergence gives 
\begin{align*}
\nabla^j\nabla^m C_{jklm} = -\frac{n-3}{n-2}  \left[ \nabla^2R_{kl}-\frac{g_{kl}\nabla^2 R}{2(n-1)}
-[\nabla_j,\nabla_k]R_l{}^j \right ]+ \frac{n-3}{2(n-1)} \nabla_k\nabla_l R.
\end{align*}
Ricci semi-symmetry and the property $R=0$ give \eqref{eq33}.
\end{proof}
\end{lem}
\begin{lem}[see \cite{ManMol2016} and \cite{ManSuh2016} theorem 6.1.]\label{3.3}
A pp-wave space-time of dimension $d+2\ge 4$ with the metric \eqref{2.6metric} satisfies $\nabla_m C_{jkl}{}^m=0$ if 
and only if $\nabla_k \psi =\lambda X_k$. 
\begin{proof}
For a pp-wave space-time we have $\nabla_j R_{kl}= (\nabla_j\psi) X_k X_l$ and $R=0$. Eq.\eqref{divC}
becomes:
\begin{align}
\nabla_m C_{jkl}{}^m=% \tfrac{n-3}{n-2} [\nabla_k R_{jl} - \nabla_j R_{kl}] = 
\frac{n-3}{n-2}  [(\nabla_k \psi )X_j- (\nabla_j\psi )X_k] X_l.
\end{align}
It follows that $\nabla_m C_{jkl}{}^m=0$ if and only if $(\nabla_k \psi )X_j= (\nabla_j\psi )X_k$, which is
equivalent to the condition $\nabla_k\psi=\lambda X_k$ for some scalar function.
\end{proof}
\end{lem}
\noindent
\begin{remark}\label{3.333}
For a pp-wave the condition $\nabla_m C_{jkl}{}^m=0$ is equivalent to $\nabla_k R_{jl}=\nabla_j R_{kl}$, i.e. to the Ricci tensor being 
a Codazzi tensor. This follows from \eqref{divC} and $R=0$.
\end{remark}
\begin{prop} \label{Prop_3.4}
Let ($\mathscr M, g$) be a $d+2\ge 4$ dimensional pp-wave space-time with the metric \eqref{2.6metric}. 
If $\psi =-\frac{1}{2}\sum_{\rho =1}^d \partial^2 H/\partial x_\rho^2$ only depends on $u$, then $\nabla^2 R_{kl}=0$.
\begin{proof}
If $\psi=\psi (u)$ we have $\nabla_k\psi = \partial_k\psi (u)= \psi'(u) X_k$ and $\nabla_m C_{jkl}{}^m=0$ by the 
previous lemma. \\Therefore $\nabla_k R_{ij}=\psi' X_kX_iX_j$ and $\nabla^2 R_{ij} = \psi''(u) X^kX_k X_iX_j=0$.
\end{proof}
\end{prop}
Galaev classified the indecomposable conformally recurrent Lorentzian manifolds \cite{Gal12}. He showed that either the 
manifold is conformally flat, or $\nabla_i  R_{jklm}=0$, or it is a pp-wave. Then he found all pp-wave potential functions $H(\vec x, u)$ such that the Weyl tensor in the metric \eqref{2.6metric} is recurrent, 
$\nabla_i C_{jklm}=\alpha_i C_{jklm}$. The potential solves the following system of $d+1$ equations:
\begin{align}
&(\alpha_\rho - \partial_\rho) \Omega_{\mu\nu} = 0 \quad (\rho=1\dots d)\\
&(\alpha_u - \partial_u) \Omega_{\mu\nu}= 0
\end{align}
where $\Omega_{\mu\nu} =  \partial_\mu\partial_\nu H-\tfrac{1}{n}\delta_{\mu\nu}\sum_\rho\partial^2_\rho H $ ($\mu,\nu=1\dots d$).\\ 
Let  $\Omega^2=\sum_{\mu\nu}(\Omega_{\mu\nu})^2$.
In an open set $\mathcal O\subset \mathscr M$ where $\Omega^2 \neq 0$ the equations give 
$\alpha_\mu=\frac{1}{2}\partial_\mu \log \Omega^2$ and $\alpha_u=\frac{1}{2}\partial_u \log \Omega^2$.
The potential for which the metric \eqref{2.6metric} is conformally recurrent turns out to be
\begin{align}
H(\vec x,u) = \sum_{\rho =1}^d x_\rho^2 [a(u)+ F(u)\lambda_\rho^2] \label{eq(3.4)}
\end{align}
with functions $a(u)$, $F(u)$ and real numbers $\lambda_1+\dots+\lambda_d=0$. Correspondingly,  
$\psi (\vec x, u) = -na(u)$ and, since it only depends on $u$, 
we conclude that for conformally recurrent pp-waves 
the divergence of the conformal tensor vanishes:
\begin{prop} \label{Prop_3.555}
On a conformally recurrent pp-wave space-time with the
metric \eqref{2.6metric}, we have $\nabla_m C_{jkl}{}^m=0$; 
moreover there exists a domain $\mathcal O$ where the recurrence covector is a gradient.
\end{prop}
The recurrence 1-form $\alpha_i$ being closed, it follows that $[\nabla_p, \nabla_q] C_{jklm}=0$, 
i.e. the manifold is Weyl semi-symmetric (see \cite{Deszcz90}). Moreover, as $[\nabla_i,\nabla_j ]R_{kl}=0$, 
the manifold is also semi-symmetric, $[\nabla_p,\nabla_q] R_{jklm}=0$.\\
Now, from Proposition \ref{Prop_3.555}, Remark \ref{3.333} and Theorem \ref{th2.2}, we have:
\begin{thm} \label{3.8}
In $n\ge 4$ any conformally recurrent pp-wave space-time is a simple conformally recurrent Lorentzian manifold.
\end{thm}
Some further properties of conformally recurrent pp-wave space-times are collected here:
\begin{prop} %3.10, 3.11
For a conformally recurrent pp-wave space-time:\\
1) the recurrence covector $\alpha_i$ is null  and collinear with the covariantly constant covector $X_i=\nabla_i u$\\
2) the recurrence covector is recurrent with closed recurrence 1-form. 
\begin{proof}
From Propositions \ref{Prop_3.555} and \ref{prop_Adati} we get 
$\alpha_iC_{jklm} +\alpha_jC_{kilm}+\alpha_k C_{ijlm}=0$ i.e. the recurrence
covector of a conformally recurrent pp-wave space-time belongs to the Olszak distribution. 
In view of Remark \ref{rem_2.11}, 
for a pp-wave space-time also $X_i$ belongs to the Olszak distribution; then, by Lemma \ref{Lem3.9}, $\alpha_i=\mu X_i$. 
Taking the covariant derivative, we see that 
$\nabla_j\alpha_i = (\nabla_j\mu)X_i  = (\nabla_j \log |\mu|) \alpha_i \equiv q_j\alpha_i $ (i.e. the
recurrence vector is itself recurrent, with closed recurrence 1-form).
\end{proof}
\end{prop}
\begin{prop}%3.12.
Let ($\mathscr M,g$) be a $d+2\ge 4$ dimensional pp-wave space-time with the metric \eqref{2.6metric}. Then 
there exists a coordinate domain $\mathcal O$ in $\mathscr M$ where $\nabla^2 C_{jklm}=0$ and $\nabla^2 R_{jklm}=0$.
\begin{proof}
From the closedness condition $\nabla_j\alpha_i = \nabla_i\alpha_j$ we infer $q_i\alpha_j=q_j\alpha_i$ so that $q_j=\rho\alpha_i$ 
and the recurrence 1-form $\alpha_i$ satisfies $\nabla_j\alpha_i=\rho\alpha_i\alpha_j$ and $\nabla^i\alpha_i=0$. Thus, since 
$\alpha^i\alpha_i=0$, we have $\alpha^i\nabla_i C_{jklm}=0$ and consequently $\nabla^i\nabla_i C_{jklm}= \nabla^i (\alpha_i C_{jklm} ) = 
(\nabla^i\alpha_i)C_{jklm}+
\alpha^i (\nabla_iC_{jklm})=0$ . Moreover we have $\nabla^i\nabla_i R_{kl}=0$ so that $\nabla^i\nabla_i R_{jklm}=0$.
\end{proof}
\end{prop}

Now we consider a $n$-dimensional Lorentzian simple conformally recurrent manifold and show that it 
is a conformally recurrent pp-wave space-time.
\begin{thm}\label{3.13}
Let ($\mathscr M,g$) be a Lorentzian simple conformally recurrent manifold of dimension $n\ge 4$: then there exists a  
coordinate domain $\mathcal O$ in $\mathscr M$ where the manifold is a pp-wave space-time. 
\begin{proof}
From Theorem \ref{th2.2} and Proposition \ref{prop2.3} we have $\nabla_m  C_{jkl}{}^m=0$ so that $\alpha^m C_{jklm}=0$ and
$\alpha_i C_{jklm}+\alpha_j C_{kilm} + \alpha_k C_{ijlm} = 0$. 
Taking the covariant derivative, we get   
$(\nabla_p\alpha_i) C_{jklm}+(\nabla_p\alpha_j) C_{kilm} + (\nabla_p\alpha_k) C_{ijlm} = 0$. 
Since in a Lorentzian manifold the Olszak distribution is one-dimensional (Lemma \ref{Lem3.9}), 
this implies proportionality, $\nabla_i \alpha_j=p_i\alpha_j $  for
some 1-form $p_j$.

Proposition \ref{Prop2.5} states that a simple conformally recurrent Lorentzian
manifold is Ricci recurrent, i.e. $\nabla_k R_{jl}=\omega_k R_{jl}$ for some 1-form $\omega_k$. Furthermore, by 
Theorem \ref{th2.2}, the Ricci tensor is Codazzi, $\nabla_k R_{jl}=\nabla_j R_{kl}$.  Then we have
\begin{align}
\omega_j R_{kl}= \omega_k R_{jl}. \label{3.5}
 \end{align}
Contraction with $g^{jl}$ gives $\omega^l R_{kl}=0$ because $R=0$ (theorem 2.3). Thus on multiplying 
\eqref{3.5} by $\omega_k$ it follows that
$(\omega^k\omega_k)R_{jl}=0$, i.e. $\omega_k$ is a null vector. 

Let $\theta^k$ be a vector such that $\theta^k\omega_k=1$. Eq.\eqref{3.5} gives
$R_{jl}= \omega_j\theta^k R_{kl}$ and, by symmetry, $\omega_j \theta^k R_{kl}= \omega_l\theta^k R_{kj}$. Thus 
$\theta^j R_{jl} = \omega_l (\theta^k\theta^j R_{kj})$ from which we finally get: 
\begin{align}
R_{ij} = \psi \omega_i\omega_j  % eq 3.7
\end{align}
where $\psi=\theta^m\theta^j R_{mj}$ is a scalar function.\\
Let $d= \psi/|\psi|$ and $d_j = \omega_j \sqrt{|\psi| }$. Then
$R_{ij} = d d_i d_j$ with $d^k d_k=0$. By Proposition \ref{Lem2.6}, equations \eqref{(2.3)} and \eqref{(2.4)} are 
recovered.
%, i.e.$d_i C_{jklm}+d_j C_{kilm}+d_k C_{ijlm}=0$ and $d_i R_{jklm}+d_j R_{kilm}+d_k R_{ijlm}=0$; 
In this way the vector $d_j$ belongs to the one-dimensional Olszak distribution: thus $d_i= \epsilon \alpha_i$ 
and the Ricci tensor takes the form $R_{jl}=\phi\alpha_j\alpha_l$.
Furthermore, from 
\begin{align}
\alpha_i R_{jklm}+\alpha_j R_{kilm}+\alpha_k R_{ijlm}=0 \label{alphaRiem}
\end{align} 
Remark \ref{rem_2.11} gives: $\alpha^m R_{jklm}=0$. Taking the covariant derivative of $\nabla_k\alpha_l=p_k\alpha_l$ and using skew symmetrisation, we obtain 
$\alpha^m R_{jklm}  = (\nabla_j p_k - \nabla_k p_j)\alpha_l = 0$. Thus the covector $p_j$ is locally a gradient, i.e. there 
exists a coordinate domain $\mathcal O$ of 
($\mathscr  M,g$) where $p_j=\nabla_j \eta $. 
The covector $\overline\alpha_j=\alpha_j e^{-\eta}$ is covariantly constant ($\nabla_k \overline\alpha_j=0$). Eq.\eqref{alphaRiem}
with $\bar\alpha_j$ is Schimming's condition \eqref{222} for the manifold to be locally a pp-wave space-time.
\end{proof}
\end{thm}

\section{Some consequences for extended theories of gravitation}
We derive some consequences for extended theories of gravitation in dimensions $n\ge 4$. 
Since for a pp-wave the scalar curvature is zero, a pp-wave solves 
the Einstein's field equations (in natural units) 
$$R_{ij}-\tfrac{1}{2}Rg_{ij}=8\pi T_{ij} $$ 
with a pure radiation source, $T_{ij}=\Phi^2 k_ik_j$, where $k_ik^i=0$ (see \cite{Stephani}, eq. 5.8).\\
Generic theories of gravitation modify Einstein's theory at short distances by expressing the action integral
with zero source as a power series 
\begin{align}
 I=&\int d^n x\, \sqrt{-g}  \Big[ R-2\Lambda_0 +\alpha R^2 +\beta R_{ij}R^{ij} \\
& +\gamma (R_{jklm} R^{jklm} -4R_{jk}R^{jk} +R^2) +\sum_{p>2}
C_p (\text{Riem, Ric, $\nabla$Riem,} \dots) \Big ] \nonumber
\end{align}
where $\alpha$, $\beta$, $\gamma, C_p$ are parameters provided by some microscopic theory such as
string theory. The quadratic part represents ``quadratic gravity" (Weyl-Eddington terms), and the terms of higher order are
contracted products of the Riemann tensor and its derivatives. Such corrections scale as powers of
$\ell_P/L$, where $\ell_P$ is Planck's length and $L$ is a typical length for the variation of the metric \cite{Zee},
and could be significant in the early evolution of the universe.\\
The field equations of the full theory are complicated; nevertheless G\"urses et al. \cite{Gur14} 
proved that for pp-wave metrics \eqref{2.6metric} the field equations with cosmological constant $\Lambda_0=0$ may be written as:
\begin{align}
[a_0 + a_1 \nabla^2 + a_2 (\nabla^2)^2 +\dots ] R_{kl}=0  \label{4.2}
\end{align}
where $a_i$ are constants depending on $\alpha,\beta,\gamma, C_p$. 
If $\nabla^2 R_{kl}=0$  the equation reduces to $a_0 R_{kl}=0$, with the vacuum solution.
In the same paper the authors noted that for pp-waves with $\nabla^2 R_{kl}=0$ the field equations may include a pure radiation source:
\begin{align}
a_0 R_{kl} =  T_{kl} \label{4.3}
\end{align}
It is thus interesting to find non-vacuum pp-wave solutions. 
According to Proposition \ref{Prop_3.4}, if $\psi=-\frac{1}{2}\sum_{\rho=1}^d \partial_\rho^2 H$ depends only on $u$,
then $\nabla^2 R_{kl}=0$. 
\begin{prop}
Let $(\mathscr M,g)$ be a $d+2\ge 4$ dimensional pp-wave space-time with the metric \eqref{2.6metric}: 
if $\sum_{\rho =1}^d \partial_\rho^2 H$ 
depends only on $u$,  then 
the metric solves the field equations of the generic theory of gravitation with pure 
radiation source. 
\end{prop}
\noindent
We give two examples where $\nabla^2 R_{kl}=0$:\\
1) the pp-wave metric is conformally recurrent, with potential
$H$ given by \eqref{eq(3.4)}. \\
2) the pp-wave metric is two-symmetric i.e. $\nabla_p \nabla_i R_{jklm}=0$. This occurs if and only 
if the potential function of the pp-wave has the form 
\begin{align}
H(\vec x, u )=\sum_{\mu,\nu=1}^d (u a_\mu \delta_{\mu\nu} + b_{\mu\nu}) x^\mu x^\nu
\end{align}
where $0\le a_1\le \dots \le a_d$ and $b_{\mu\nu}=b_{\nu\mu}$  are real numbers (see \cite{Ale11,Bla10,Fer12}). One evaluates
$\psi(u)= -\sum_{\mu=1}^d (ua_\mu+b_{\mu\mu})$. Thus for 
two-symmetric pp-waves space-times the divergence of the conformal tensor vanishes and, as one may 
directly calculate, $\nabla^2 R_{kl}=0$. 
%
%%%%%%%%%%%%%%%%%%%%%%%%%%%%%%%%%%%%
%

\end{document}